\documentclass[a4paper,11pt]{article}
\usepackage{amssymb,palatino}
\newtheorem{theorem}{Theorem}

\newcommand{\QED}{\begin{flushright} $\Box$ \end{flushright}}
\begin{document}

\title{Polynomial evaluation over finite fields:\\ new algorithms and complexity bounds\thanks{This is an extended version of the paper 'Efficient evaluation of polynomials over finite fields' presented at the 2011 Australian Communications Theory Workshop, Melbourne, Victoria, January 31 - Feburary 3, 2011.}}

\author{Michele Elia\thanks{Politecnico di Torino, Italy}, Joachim Rosenthal\thanks{www.math.uzh.ch/aa}, Davide Schipani\thanks{University of Zurich, Switzerland}~~  
}

\maketitle

\thispagestyle{empty}

\begin{abstract}
\noindent
An efficient evaluation method is described for polynomials in finite fields. 
Its complexity is shown to be lower than that of
 standard techniques, when the degree of the polynomial is large enough compared to the field characteristic. 
Specifically, if $n$ is the degree of the polynomiaI, the asymptotic complexity is shown to be $O(\sqrt{n})$, versus
 $O(n)$ of classical algorithms.
Applications to the syndrome computation in the decoding of Reed-Solomon codes are highlighted.
\end{abstract}

\vspace{2mm}

\vspace{2mm}

\vspace{8mm}

\section{Introduction}
The direct evaluation of a polynomial $P(x)=a_n x^n+a_{n-1} x^{n-1}\cdots+a_0$ of
 degree $n$ over a ring or a field in a point $\alpha$ may be performed
 computing the $n$ powers $\alpha^i$ recursively as $\eta_{i+1} =\alpha \eta_i$, for $i=1, \ldots , n-1$,
 starting with $\eta_1=\alpha$,  obtaining $P(\alpha)$ as
$$ P(\alpha)= a_0+ a_1 \eta_1 + a_2 \eta_2 + \cdots + a_n \eta_n  ~~. $$ 
This method requires $2n-1$ multiplications and $n$ additions. 
However, Horner's rule (e.g. \cite
{knuth2}), which has become a standard, is more efficient and
 computes the value $P(\alpha)$ iteratively as
$$
P(\alpha)=  (\cdots((a_n\alpha+a_{n-1})\alpha+a_{n-2})\alpha+\cdots)\alpha+a_1)\alpha
+a_0 ~~.
$$
This method requires $n$ multiplications and $n$ additions.
%
In particular scenarios, for example when the number of possible values of the coefficients is finite,
more advantageous procedures can be used, as it will be shown in this document.

We point out that what is usually considered in the literature to establish upper and lower bounds to the
 minimum number of both "scalar" and "nonscalar" multiplications refers, sometimes implicitly, to
 polynomials with coefficients taken from an infinite set, e.g. fields of characteristic zero, or
 algebraically closed fields.  
In fact, in \cite{borodin,pan,winograd}, Horner's rule is proved to 
 be optimal assuming that the field of coefficients is infinite; instead, we show that this is not the case
 if the coefficients belong to a finite field.
Furthermore, in \cite{paterson}, restricting the field of coefficients to the rational field, and converting 
 multiplications by integers into iterated sums (therefore scalar multiplications are not counted in that model), it is shown that the number of required multiplications   
 is less than that required by Horner's rule, although the number of sums can grow unboundedly. 

In the following we describe a method to evaluate polynomials with
coefficients over a finite field $\mathbb F_{p^s}$, and estimate its complexity in terms of field multiplications
 and sums. 
However, as is customary, we only focus on the number of multiplications,
 that are more expensive operations than additions: in $\mathbb F_{2^m}$, for example, the cost
 of an addition is $O(m)$ in space and $1$ clock in time, while the cost of a
 multiplication is $O(m^2)$ in space and $O(\log_2 m)$ in time (\cite{elia}).
Clearly, field multiplication by  look-up tables may be faster, but this approach is only possible for small values of $m$. 
We also keep track of the number of additions, so as to verify that a reduction in the number of
 multiplications does not bring with it an exorbitant increase in the number of additions.  \\
Our approach exploits the Frobenius automorphism and its group
properties, therefore we call it "polynomial automorphic evaluation".

The next Section describes the principle of the algorithm,
 with two different methods, referring to the evaluation
 in a point of $\mathbb F_{p^m}$ of a polynomial with coefficients in the prime field $\mathbb F_p$. 
The complexity is carefully estimated in order to make the comparisons self-evident.
Section 3 concerns the evaluation in $\mathbb F_{p^m}$ of polynomials with coefficients in
 $\mathbb F_{p^s}$, for any $s>1$ dividing $m$: different approaches will be described
 and their complexity compared.
Section 4 includes examples concerning the syndrome computation in the algebraic decoding of error-correcting codes
 (cf. also \cite{schip}), and some final remarks. 

\section{Polynomial automorphic evaluation: basic principle}
Consider a polynomial $P(x)$ of degree $n >p$ over a prime field $\mathbb F_p$,
 and let $\alpha$ be an element of $\mathbb F_{p^m}$. 
We write $P(x)$ as a sum of $p$ polynomials
\begin{equation}
  \label{basicdec}
P(x)= P_{1,0}(x^p)+ x P_{1,1}(x^p)\cdots +x^{p-1} P_{1,p-1}(x^p) ~~,
\end{equation}
where $P_0(x^p)$ collects the powers of $x$ with exponent a multiple
of $p$ and in general $x^{i} P_{i}(x^p)$ collects the powers of the form
$x^{ap+i}$, with $a\in\mathbb{N}$ and $0\leq i\leq p-1$. \\

\paragraph{First method.}

If $\sigma$ is the Frobenius automorphism of $\mathbb F_{p^m}$ mapping $\gamma$ to
$\gamma^p$, which leaves invariant the elements of $\mathbb F_{p}$, we write the
 expression above as
$$
P_{1,0}(\sigma(x))+ x P_{1,1}(\sigma(x))+ \cdots +x^{p-1} P_{1,p-1}(\sigma(x)) ~~,
$$
where $P_{1,i}(y)$, $i=0, \ldots , p-1$, are polynomials of degree $\lfloor \frac{n}{p}\rfloor$ at most.
Then we may evaluate these $p$ polynomials in the same point $\sigma(\alpha)$, and obtain $P(\alpha)$ as
 the linear combination
$$ 
P_{1,0}(\sigma(\alpha))+ \alpha P_{1,1}(\sigma(\alpha))\cdots + 
     \alpha^{p-1} P_{1,p-1}(\sigma(\alpha)) ~~.   $$
A possible strategy is now to evaluate recursively the powers $\alpha^j$ for $j$ from $2$ up to $p$,
 and $\sigma(\alpha)^j$ for  $j$ from $2$ up to $\lfloor \frac{n}{p}\rfloor$,
 compute the $p$ numbers $P_{1,i}(\sigma(\alpha))$,
 $i=0, \ldots , p-1$, using $n$ sums and at most $\lfloor \frac{n}{p}\rfloor (p-2)$ products (the powers of
 $\sigma(\alpha)$ times their possible coefficients; the multiplications by $0$ and $1$ are not counted), and obtain $P(\alpha)$ with $p-1$ products and $p-1$
 additions. 
The total number $M_p(n)$ of multiplications is 
$$ M_p(n)= p-1+ \lfloor \frac{n}{p}\rfloor -1 + (p-1)+ \lfloor \frac{n}{p}\rfloor  (p-2)
      = 2p-3 +\lfloor \frac{n}{p}\rfloor  (p-1) ~~. $$
Then this procedure is more efficient compared to 
 Horner's rule as far as $M_p(n) < n$. For example, if
 $p=3$ and $n=10$ we have
  $M_3(10)= 9 < 10$, and for every $n >10$ the outlined method is always more efficient.
  More in general the condition is certainly satisfied whenever $n > 2 p^2-3p$, as it can be
  verified by considering $n$ written in base $p$.  \\
Let us see an example in detail, for the sake of clarity, in the case $p=3$ and $n=10$. Suppose we want to evaluate the polynomial $f(x)=1+2x+x^2+2x^4+x^5+x^6+2x^8+x^{10}$
in some element $\alpha \in \mathbb F_{3^m}$. Writing $f(x)$ as in equation (\ref{basicdec}) 
$$
f(x)=1+x^6+x(2+2x^3+x^9)+x^2(1+x^3+2x^6),
$$
 we see that it is sufficient to compute $\alpha^2$, $\alpha^3$, $\alpha^6$, $\alpha^9$, then $2\alpha^3$, $2\alpha^6$, $2\alpha^9$ (all possible coefficients needed to evaluate the three sub-polymonials), and lastly the two products by $\alpha$ and $\alpha^2$ in front of the brackets, for a total of $9$ multiplications. Note that actually $2\alpha^9$ is not needed for this particular example, but in general we always suppose to have a worst case situation.
Clearly $\alpha$ should belong to $\mathbb F_{3^m}$ for some $m$ such that $3^m>n$, so that the powers of $\alpha$ up to the exponent $n$ are all different. Note, in particular, that if both the coefficients and the evaluation point are in $\mathbb F_p$, then the polynomial has degree at most $p-1$, and our methods cannot be applied.

However, the above mechanism can be iterated, and the point is to find the number of steps or iterations yielding the maximum gain. In fact we can prove the following:
\begin{theorem}
 Let $L_{opt}$ be the number of steps of this method yielding the minimum number of products, $G_1(p,n,L_{opt})$, required to evaluate a polynomial of degree $n$ with coefficients in $\mathbb F_p$. Then $L_{opt}$ is either 
 the integer which is nearest to $\log_p  \sqrt{n(p-1)}$, 
 or this integer minus $1$, and asymptotically we have:
$$
G_1(p,n,L_{opt}) \approx 2\sqrt{n (p-1)}~~.
$$
\end{theorem}
\noindent
{\sc Proof}.

At step $i$, the number of polynomials at step $i-1$ is multiplied by $p$ since each polynomial
 $P_{i-1,h}(x)$ is partitioned into $p$ sub-polynomials $P_{i,j+ph}(x)$ , $j$ varies between $0$ and $p-1$,
 of degree roughly equal to the degree of $P_{i-1,h}(x)$ divided by $p$, that is of degree 
 $\lfloor \frac{n}{p^i}\rfloor$; the number of these polynomials is $p^i$. \\
After $L$ steps we need to evaluate $p^L$ polynomials of degree nearly $\frac{n}{p^L}$, then
 $P(\alpha)$ is reconstructed performing back the linear combinations with the polynomials $P_{i,h}(x)$
  substituted by the corresponding values $P_{i,h}(\alpha)$.
 The total cost of the procedure, in terms of multiplications and additions, is composed of the
 following partial costs
\begin{itemize}
  \item Evaluation of $p$ powers of $\alpha$, this step also produces $\sigma(\alpha)=\alpha^p$,
    and requires $p-1$ products.
  \item Evaluation of $(\sigma^i(\alpha))^j$, $i=1,\dots,L-1$, $j=2,\dots,p$; this step 
    also produces $\sigma^L(\alpha)$, and requires $(p-1)(L-1)$ products.
  \item Evaluation of $\lfloor \frac{n}{p^L} \rfloor$ powers of $\sigma^L(\alpha)$, this step requires
     $\lfloor \frac{n}{p^L} \rfloor -1$ products. 
  \item Evaluation of $p^L$ polynomials $P_{L,j}(x)$, of degree at most $\lfloor \frac{n}{p^L} \rfloor$,
    at the same point $\sigma^L(\alpha)$, this step requires $n$ additions and 
    $\lfloor \frac{n}{p^L} \rfloor(p-2)$ products at most.
  \item Computation of $p-1+(p^2-p)+\cdots + p^L-p^{L-1}=p^{L}-1$  multiplications by powers of
    $\sigma^i(\alpha)$, ($i=0, \dots , L-1$).
  \item Computation of $p-1+(p^2-p)+\cdots + p^L-p^{L-1}=p^{L}-1$ additions.
\end{itemize}

\noindent
The total number of
 products as a function of $n$, $p$ and $L$ is then
$$  G_1(p,n,L) = \lfloor \frac{n}{p^L} \rfloor(p-1)+ L(p-1)+ p^{L}-2   ~~, $$
which should be minimized with respect to $L$. The values of $L$ that correspond to local minima
are specified by the conditions
\begin{equation}
   \label{inequal}
   G_1(p,n,L) \leq G_1(p,n,L-1) ~~~~\mbox{and}~~~~ G_1(p,n,L) \leq G_1(p,n,L+1)  ~~,
\end{equation}   
which can be explicitly written in the forms 
$$   \lfloor \frac{n}{p^L} \rfloor +p^{L-1} \leq \lfloor \frac{n}{p^{L-1}} \rfloor -1 ~~~~\mbox{and}~~~~
     \lfloor \frac{n}{p^L} \rfloor -p^{L}   \leq \lfloor \frac{n}{p^{L+1}} \rfloor +1   ~~.   $$
Let $\{x\}$ denote the fractional part of $x$, then $\lfloor x \rfloor= x-\{x\}$, thus the last
 inequalities can be written as
$$  1   +  \{ \frac{n}{p^{L-1}} \}-\{ \frac{n}{p^L} \}  \leq \frac{n}{p^{L-1}}-\frac{n}{p^L} -p^{L-1}
       ~~~~\mbox{and}~~~~
      \frac{n}{p^L}- \frac{n}{p^{L+1}}  -p^{L}   \leq  1+\{ \frac{n}{p^L} \}-\{ \frac{n}{p^{L+1}} \}   ~~.   $$ 
Since $\{x\}$ is a number less than $1$, these inequalities can be relaxed to
$$ 0 < \frac{n}{p^{L-1}}-\frac{n}{p^L} -p^{L-1}
       ~~~~\mbox{and}~~~~
      \frac{n}{p^L}- \frac{n}{p^{L+1}}  -p^{L}   < 2   ~~,   $$
which imply
$$ p^{2L}  < n(p-1) p
       ~~~~\mbox{and}~~~~
      n(p-1) +p  < p^{2L+1}+ 2 p^{L+1} + p =p(p^L+1)^2 ~~.   $$
Thus, we have the chain of inequalities
$$  \frac{1}{\sqrt{p}} \sqrt{n(p-1) +p} -1 < p^{L} < \sqrt{p}  \sqrt{n(p-1)} ~~,  $$
and taking the logarithm to base $p$ we have
\begin{equation}
  \label{optvalue1}
 -\log_p\left( \sqrt{1 +\frac{p}{n(p-1)}}+\sqrt{\frac{p}{n(p-1)}} \right) - \frac{1}{2}+ \log_p  \sqrt{n(p-1)}  < L < \log_p  \sqrt{n(p-1)} + \frac{1}{2}~~,
\end{equation} 
which shows that at most two values of $L$ satisfy the conditions for a minimum, because $L$ is constrained
 to be in an interval of amplitude $1+\epsilon$, with 
 $\epsilon=\log_p\left( \sqrt{1 +\frac{p}{n(p-1)}}+\sqrt{\frac{p}{n(p-1)}} \right)< 1$, around the point of
 coordinate $\log_p  \sqrt{n(p-1)}$. Therefore, the optimal value $L_{opt}$ is either 
 the integer which is nearest to $\log_p  \sqrt{n(p-1)}$, 
 or this integer minus $1$. 
Hence, we have the very good asymptotic estimation $ L_{opt} \approx  \log_p \sqrt{n (p-1)}$, and correspondingly a very good asymptotic estimation for $G_1(p,n,L_{opt})$, that is 
$$ G_1(p,n,L_{opt}) \approx 2\sqrt{n (p-1)}~~.$$
\QED

\paragraph{Second method.}

We describe here another approach exploiting the Frobenius automorphism in a different way; although it will appear to be asymptotically less efficient than the above method, it may be useful in particular situations, as shown in Section 4.
 
\noindent Since the coefficients are in $\mathbb F_{p}$,
$$
P(x)= P_{1,0}(x^p)+ x P_{1,1}(x^p)\cdots +x^{p-1} P_{1,p-1}(x^p) ~~
$$
can be written as
$$
P_{1,0}(x)^p+ x P_{1,1}(x)^p\cdots +x^{p-1} P_{1,p-1}(x)^p ~~,
$$
where $P_{1,i}(x)$, $i=0, \ldots , p-1$, are polynomials of degree $\lfloor \frac{n}{p} \rfloor$ at most.
Then we may evaluate these $p$ polynomial in the same point $\alpha$, and obtain $P(\alpha)$ as
 the linear combination
$$ 
P_{1,0}(\alpha)^p+ \alpha P_{1,1}(\alpha)^p\cdots +\alpha^{p-1} P_{1,p-1}(\alpha)^p ~~.
$$
A possible strategy is to evaluate recursively the powers $\alpha^j$ for
 $j=2,\ldots, \lfloor\frac{n}{p}\rfloor $, 
  compute the $p$ numbers $P_{1,i}(\alpha)$, $i=0, \ldots , p-1$,
using sums and at most $\lfloor\frac{n}{p}\rfloor(p-2)$ products (the powers of $\alpha$ times their possible coefficients), and obtain $P(\alpha)$ with $p$ $p$-th powers, $p-1$ products and $p-1$ additions.
The total number of multiplications is $\lfloor\frac{n}{p}\rfloor -1 + (p-1)+ pc_p+ \lfloor\frac{n}{p}\rfloor(p-2)$, where $c_p$
  denotes the number of products required by a $p$-th power (so $c_2=1$ and
  $c_p\leq 2\lfloor\log_2 p\rfloor$). 
The mechanism may be iterated:
after $L$ steps we need to evaluate $p^L$ polynomials of degree nearly $\frac{n}{p^L}$, then
 $P(\alpha)$ is reconstructed performing back the linear combinations with the $p$-powers of the polynomials
 $P_{i,h}(x)$ substituted by the corresponding values $P_{i,h}(\alpha)$.
\begin{theorem}
 Let $L_{opt}$ be the number of steps of this method yielding the minimum number of products, $G_2(p,n,L_{opt})$, required to evaluate a polynomial of degree $n$ with coefficients in $\mathbb F_p$. Then $L_{opt}$ lies in an interval around $\log_p \sqrt{\frac{n(p-1)^2}{pc_p+p-1}}$ of length at most $2$, and asymptotically we have:
$$
G_2(p,n,L_{opt}) \approx 2\sqrt{n(pc_p+p-1)}~~.
$$
\end{theorem}
\noindent
{\sc Proof}.

 The total cost
 of the procedure, in terms of multiplications and additions, is composed of the
 following partial costs
\begin{itemize}
  \item Evaluation of $\lfloor \frac{n}{p^L} \rfloor$ powers of $\alpha$.
  \item Evaluation of $p^L$ polynomials $P_{L,j}(x)$, of degree at most $\lfloor \frac{n}{p^L} \rfloor$,
    at the same point $\alpha$, this step requires $n$ additions and $\lfloor \frac{n}{p^L} \rfloor(p-2)$ products.
  \item Computation of $p+p^2+ \cdots + p^L=\frac{p^{L+1}-p}{p-1}$ $p$-th powers.
  \item Computation of $p-1+(p^2-p)+\cdots + p^L-p^{L-1}=p^{L}-1$  multiplications by powers of $\alpha$.
  \item Computation of $p-1+(p^2-p)+\cdots + p^L-p^{L-1}=p^{L}-1$ additions.
\end{itemize}

\noindent
Then the total number of
 products as a function of $n$, $p$ and $L$ is
$$  G_2(p,n,L) = \lfloor \frac{n}{p^L} \rfloor -1 + \frac{p^{L+1}-p}{p-1} c_p + 
               (p^{L}-1)+\lfloor \frac{n}{p^L} \rfloor(p-2) ~~, $$
which should be minimized with respect to $L$. The optimal value of $L$ is obtained by 
 conditions analogous to (\ref{inequal}) and arguing as above we find that this optimal value must be included in
 a very small interval. 

Setting $y= 4 n (p c_p +p-1) \frac{1}{p}$, the optimal value for $L$ turns out to be included into an interval around
     $L_1=\log_p \sqrt{\frac{n(p-1)^2}{pc_p+p-1}}$ of extremes
$$  L_1 -\frac{1}{2} -\log_p \left(\sqrt{1+\frac{1}{y}} +\sqrt{\frac{1}{y}}\right) ~~~~\mbox{and} ~~~~
  L_1+\frac{1}{2}+\log_p \left(\sqrt{1+\frac{1}{y}} +\sqrt{\frac{1}{y}}\right)   ~~, $$
which restricts the choice of $L_{opt}$ to at most two values. 
Hence, we have the very good asymptotic estimation $ L_{opt} \approx  \log_p \sqrt{\frac{n(p-1)^2}{pc_p+p-1}}$, and correspondingly a very good asymptotic estimation for $G_2(p,n,L_{opt})$, that is 

\begin{equation}
    G_2(p,n,L_{opt}) \approx 2\sqrt{n(pc_p+p-1)}  ~~.  
\end{equation}
\QED

\subsection{$p=2$}
The prime $2$ is particularly interesting because of its occurrence in many practical applications, for example in error correction coding. In this setting an important issue is the computation of syndromes for a binary code (\cite{sloane}), 
where it is usually needed to evaluate a polynomial in several powers of a particular value, so that
an additional advantage of the proposed method may be the possibility of
precomputing the powers of $\alpha$. \\
A polynomial $P(x)$ over the binary field is simply decomposed into a sum of two polynomials
 by collecting odd and even powers of $x$ as
$$  P(x) = P_{1,0}(x^2)+ x P_{1,1}(x^2)= P_{1,0}(x)^2 + x P_{1,1}(x)^2  ~~. $$
The mechanism is then the same as for odd $p$ with a few simplifications. The main point is that we do not need to multiply with the coefficients, which are either $0$ or $1$, so only sums are finally involved when evaluating the polynomials.

\noindent

And to evaluate $2^L$ polynomials at the same point $\alpha$ we would need
  to evaluate the powers $\alpha^j$ for $j=2, \ldots, \lfloor\frac{n}{2^L}\rfloor$, and then obtain
 each $P_{Lj}(\alpha)$ by adding those powers corresponding to non-zero coefficients;
 the number of additions per each polynomial is nearly $\frac{n}{2^L}$, then the
 total number of additions is not more than $n$. 
But the actual number of additions is much smaller if sums of equal terms can be reused,
 and it is upper bounded by $O(\frac{n}{ln(n)})$. 
This bound is a consequence of the fact that in order to evaluate $2^L$ polynomials of degree
$h=\lfloor \frac{n}{2^L} \rfloor$ at the same point $\alpha$, we have to compute $2^L$ sums of the form
$$ \alpha^{i_1} + \cdots + \alpha^{i_m},~~m\leq h $$
 having at disposal the $h$ powers $\alpha^i$. We can then think of a 
 $2^L \times \lfloor \frac{n}{2^L} \rfloor$ binary matrix to be multiplied
 by a vector of powers of $\alpha$, and
 assuming $2^L \approx \frac{n}{2^L}$ (as follows from the estimation of the minimum discussed above), we may consider the matrix
 to be square and apply \cite[Theorem 2]{joachim}. 

\section{ Automorphic evaluation of polynomials over extended fields}

This section considers the evaluation in $\alpha$, an element of $\mathbb F_{p^m}$,
 of polynomials $P(x)$ of degree $n$ over $\mathbb F_{p^s}$, a subfield of $\mathbb F_{p^m}$ larger than
 $\mathbb F_{p}$, thus $s>1$ and $s|m$. There are two ways to face the problem, 
one way is more direct, the second way exploits the Frobenius automorphism.  

\paragraph{First method.}

Let $\beta$ be a generator of a polynomial basis of $\mathbb F_{p^s}$, i.e. $\beta$ is a root
 of an irreducible $s$-degree polynomial over $\mathbb F_{p}$, expressed as an element of $\mathbb F_{p^m}$, then $P(x)$ can be written as
\begin{equation}
   \label{basextfield}
   P(x)= P_0(x)+\beta P_1(x) + \beta^2 P_2(x) + \cdots + \beta^{s-1} P_{s-1}(x) ~~, 
\end{equation}    
 where $P_i(x)$, $i=0, \ldots, s-1$, are polynomials over $\mathbb F_{p}$ (cf. also \cite{sarwate}). 
Then $P(\alpha)$ can be obtained as a linear combination of the $s$ numbers $P_i(\alpha)$.
Thus the problem of evaluating $P(\alpha)$ is reduced to the problem of evaluating $s$ polynomials 
 $P_i(x)$ with $p$-ary coefficients followed by the computation of $s-1$ products and $s-1$ sums
 in $\mathbb F_{p^m}$. 

\noindent
We can state then the following:
\begin{theorem}
The minimum number of products required to evaluate a polynomial of degree $n$ with coefficients in $\mathbb F_{p^s}$ is upper bounded by $2s(\sqrt{n(p-1)}+\frac{1}{2})$.

\end{theorem}
\noindent
{\sc Proof}.
The upper bound is a consequence of Theorem 1 and the comments following equation (\ref{basextfield}).
\QED
The total complexity grows asymptotically as $2s   \sqrt{n(p-1)}$, so that a general upper bound (possibly tight)
 for the number of multiplications that are sufficient to compute $P(\alpha)$, when $P(x)$ has
 coefficients in any subfield of $\mathbb F_{p^m}$, is then $ 2m   \sqrt{n(p-1)}$.

\vspace{5mm}

\paragraph{Second method.}

This consists in generalizing the basic principle directly. We will show the following:
\begin{theorem}
$G_1(p^s,n,L_{opt})\approx 2   \sqrt{n(p^s-1)}$ and $G_2(p^s,n,L_{opt})\approx 2\sqrt{n(p^s-1)}\sqrt{1+c_{p^{s-1}}+c_p\frac{p}{p-1}}$.

\end{theorem}
\noindent
{\sc Proof}.

\noindent As for the first description, the point now is that there are $p^s-1$ possible coefficients to be multiplied, so that we get an asymptotic complexity of $G_1(p^s,n,L_{opt})\approx 2   \sqrt{n(p^s-1)}$.

\noindent Considering the second variant, $P(x)=P_{1,0}(x^p)+ x P_{1,1}(x^p)\cdots +x^{p-1} P_{1,p-1}(x^p)$ is now not directly decomposable into a sum of powers of the polynomials $P_i(x)$ since the Frobenius automorphism $\sigma$ alters their
 coefficients. However, we can write  (\ref{basicdec}) as
$$
P_{1,0}^{-1}(x)^p+ x P_{1,1}^{-1}(x)^p\cdots +x^{p-1} P_{1,p-1}^{-1}(x)^p ~~,
$$
where $P_{1,i}^{-1}(x)$ stands for the polynomial obtained from $P_{1,i}(x)$
by substituting its coefficients with their transforms through
$\sigma^{-1}$ (and if we iterate this for $k$ times we would consider $\sigma^{-k}$). 
Notice that the polynomials $P_{1,i}^{-1}(x)$ have degree at most $n_i=\frac{n-i}{p}$, and are
 obtained by computing a total of $n$ automorphisms $\sigma^{-1}$. However, in order to compute
 the $p$ numbers $P_{1,i}^{-1}(\alpha)$, $i=0, \ldots, p-1$, it is not necessary to compute
 the total number of $n$ inverse automorphisms observing that
$$  P_{1,i}^{-1}(\alpha) = \sum_{j=0}^{n_i} \sigma^{-1}(c_j) \alpha^j =
           \sigma^{-1}(\sum_{j=0}^{n_i}c_j \sigma(\alpha^j)), $$
where $c_j$, $j=1,\ldots,n_i$, are the coefficients of $P_{1,i}(x)$.
It is then sufficient to first evaluate $\sigma(\alpha)$, compute then $P_{1,i}(\sigma(\alpha))$ and
 finally apply $\sigma^{-1}$. This procedure requires the application of only $p$ automorphisms
 $\sigma^{-1}$ instead of $n$. 

If we perform $L$ steps, we need to apply $\sigma^{-L}$ a number of times not greater
  than $p^L$. Notice also that what interests us in $\sigma^L$ is $L$
  modulo $s$ because $\sigma^s$ is the identity automorphism in $\mathbb F_{p^s}$, the field of the coefficients.
The number of multiplications to be minimized becomes:
$$
  G_2(p^s,n,L)= c_p \frac{p^{L+1}-p}{p-1}+p^{L}-1+
  c_{p^{s-1}}p^L+\lfloor \frac{n}{p^L} \rfloor(p^s-1) ~~,
$$
where 
the automorphism $\sigma^L$ counts like a power with exponent $p^K$, with $K= L \bmod s \leq s-1$.
The optimal value of $L$ is obtained by analogues of conditions (\ref{inequal}) and arguing as above
 we find that this optimal value must be included in a very small interval. 

Setting $y=  \frac{4 n (p-1) (p c_p +p-1+c_{p^{s-1}}(p-1))}{p(p^s-1)}$, the optimal value for $L$
 is included into an interval around
     $L_2=\log_p \sqrt{\frac{n(p-1)(p^s-1)}{p c_p +p-1+c_{p^{s-1}}(p-1)}}$ of extremes
\begin{equation}
 \label{optpointss}
  L_2 -\frac{1}{2} -\log_p \left(\sqrt{1+\frac{1}{y}} +\sqrt{\frac{1}{y}}\right) ~~~~\mbox{and} ~~~~
  L_2+\frac{1}{2}+\log_p \left(\sqrt{1+\frac{1}{y}} +\sqrt{\frac{1}{y}}\right)   ~~, 
\end{equation}
which restricts the choice of $L_{opt}$ to at most two values.
Hence, we have the very good asymptotic estimation 
$ L_{opt} \approx  \log_p \sqrt{\frac{n(p-1)(p^s-1)}{p c_p +p-1+c_{p^{s-1}}(p-1)}}$, and correspondingly 

$$
  \label{optvalue}
  G_2(p^s,n,L_{opt})\approx 2\sqrt{n(p^s-1)}\sqrt{1+c_{p^{s-1}}+c_p\frac{p}{p-1}} ~~.
$$
\QED

\section{Examples and conclusions}
In some circumstances, for example when $s\approx m\approx \log_p n$, %
the optimal $L$ and the consequent estimated computational cost may obscure the advantages
 of the new approach, suggesting the practical use of standard techniques.
However, this might not be always a good strategy, as shown by the following example
 borrowed from the error correcting codes. \\ 
 Let us consider the Reed-Solomon codes that are used in any CD rom, or the famous
 Reed-Solomon code $[255,223,33]$ over $\mathbb F_{2^8}$ used by NASA (\cite{wicker2}):
 in such applications an efficient evaluation of polynomials over $\mathbb F_{2^m}$ in points of the
 same field is of the greatest interest (see also \cite{schip}).

What we now intend to show is that in particular scenarios the proposed methods allow additional cost reductions that can be
 obtained by a clever choice of the parameters, for example choosing $L$ as a factor of $m$ that is close to the optimal value previously found
and employing some
 other strategies as explained below.

 The idea will be illustrated considering the computation of the syndromes needed in the decoding
 of the above mentioned Reed-Solomon code. We will only show how to obtain the $32$
 syndromes; from that point onwards decoding may employ the standard
 Berlekamp-Massey algorithm, the Chien search to locate errors,
 and the Forney algorithm to compute the error
 magnitudes~(\cite{blahut}).

 Let $r(x)=\sum_{i=0}^{254} r_i x^i$, $r_i \in \mathbb F_{2^8}$, be a received
 code word of the Reed-Solomon code $[255,223,33]$ generated by the
 polynomial $g(x)=\prod_{i=1}^{32}(x-\alpha^i)$, with $\alpha$ a
 primitive element of $\mathbb F_{2^8}$, i.e. a root of $x^8+x^5+x^3+x+1$. The
 aim is to evaluate the syndromes $S_j=r(\alpha^j)$, $j=1, \ldots ,
 32$.

 A possible approach is as follows.  The power
 $\beta=\alpha^{17}$ is a primitive element of the subfield $\mathbb F_{2^4}$,
 it is a root of the polynomial $x^4+x^3+1$, and has trace $1$ in
 $\mathbb F_{2^4}$. Therefore, a root $\gamma$ of $z^2+z+\beta$ is not in
 $\mathbb F_{2^4}$ (see \cite[Corollary 3.79, p.118]{lidl}), but it is an
 element of $\mathbb F_{2^8}$, and every element of  $\mathbb F_{2^8}$ can be written
 as $a+b \gamma$ with $a,b \in \mathbb F_{2^4}$.
 Consequently, we can write $r(x)=r_1(x)+\gamma r_2(x)$ as a sum of
 two polynomials over $\mathbb F_{2^4}$, evaluate each $r_i(x)$ in the roots
 $\alpha^j$ of $g(x)$, and obtain each syndrome
 $S_j=r(\alpha^j)=r_1(\alpha^j)+\gamma r_2(\alpha^j)$ with $1$
 multiplication and $1$ sum.

 Now, we choose to adopt our second variant which turns out to be very well-suited since we will actually avoid to compute any automorphism. If $p(x)$ is either $r_1(x)$ or
 $r_2(x)$, in order to evaluate $p(\alpha^j)$ we must consider the
 decomposition
$$
p(x) = (\sigma^{-1}(p_0)+\sigma^{-1}(p_2)x+\cdots +\sigma^{-1}(p_{254})x^{127})^2 
+ x (\sigma^{-1}(p_1)+ \sigma^{-1}(p_3)x+\cdots +\sigma^{-1}(p_{253})x^{126})^2~~. 
$$
Now, each of the two parts can be decomposed again into the sum of two
polynomials of degree at most $63$, for instance
$$
  \sigma^{-1}(p_0)+\sigma^{-1}(p_2)x+\cdots +\sigma^{-1}(p_{254})x^{127} = (\sigma^{-2}(p_0)+\sigma^{-2}(p_4)x+\cdots
  +\sigma^{-2}(p_{252})x^{63})^2+
$$
$$
   x (\sigma^{-2}(p_2)+\sigma^{-2}(p_6)x+\cdots +\sigma^{-2}(p_{254})x^{63})^2
$$
and at this stage we have four polynomials to be evaluated.  The next
two steps double the number of polynomials and halve their degree; one polynomial per each stage is given here as an example
$$
  \sigma^{-2}(p_0)+\sigma^{-2}(p_4)x+\cdots +\sigma^{-2}(p_{252})x^{63}
  = (\sigma^{-3}(p_0)+\sigma^{-3}(p_8)x+\cdots +\sigma^{-3}(p_{248})x^{31})^2 +
$$
$$
   x(\sigma^{-3}(p_4)+\sigma^{-3}(p_{12})x+\cdots +\sigma^{-3}(p_{252})x^{31})^2
$$
$$
\sigma^{-3}(p_0)+\sigma^{-3}(p_8)x+\cdots +\sigma^{-3}(p_{248})x^{31}
 =(\sigma^{-4}(p_0)+\sigma^{-4}(p_{16})x+\cdots +\sigma^{-4}(p_{240})x^{15} )^2 +
$$
$$
x(\sigma^{-4}(p_8)+\sigma^{-4}(p_{24})x+\cdots +\sigma^{-4}(p_{248})x^{15})^2  
$$

Since we choose to halt the decomposition at this stage (notice that $L=4$ is a putative optimal value
 given by (\ref{optpointss})), we must
evaluate $16$ polynomials of degree at most $15$ with coefficients in
$\mathbb F_{2^4}$. 
We do not need to compute $\sigma^{-4}$ on the coefficients, as
$\sigma^{-4}(p_i)=p_i$, since the coefficients are in
$\mathbb F_{2^4}$ and any element $\beta$ in this field satisfies the condition
$\beta^{2^4}=\beta$.


We remark that up to know we have only indicated how to partition the original polynomial. This task does not require any computation, it just defines in which order to read the coefficients of the original polynomial. 

Now, let $K$ be the number of code words to be decoded. We compute only once the following field elements:
\begin{itemize}
\item $\alpha^i$, $i= 2, \ldots , 254$ and this requires $253$
  multiplications;
\item $\alpha^i \cdot \beta^j$ for $i=0,\ldots ,254$ and $j=1, \ldots,
  14$, which requires $255 \cdot 14=3570$ multiplications.
\end{itemize}
Then only sums (that can be performed in parallel) are required to
evaluate $16$ polynomials of degree $15$ for each $\alpha^j$, $j=1
\ldots , 32$. Once we have the values of these polynomials, in order
to reconstruct each of $r_1(\alpha^j)$ and $r_2(\alpha^j)$, we need
\begin{itemize}
\item $16+8+4+2$ squares
\item $8+4+2+1$ multiplications (and the same number of sums).
\end{itemize}
Summing up, every $r(\alpha^j)=r_1(\alpha^j)+\gamma r_2(\alpha^j)$ is
obtained with $2 \cdot 45+1=91$ multiplications.  Then the total cost
of the computation of $32$ syndromes drops down from $31+32 \cdot
254=8159$ with Horner's rule to $32 \cdot 91+3570+253 =6735$. Since we
have $K$ code words the total cost drops from $31+8128 \cdot K$ to $3823+
2912 \cdot K$, with two further advantages:

- many operations can be parallelized, further
increasing the speed;

- the multiplications can be performed in $\mathbb F_{2^4}$ instead of
$\mathbb F_{2^8}$, if we write $\alpha^j=a_j+\gamma b_j$; this might increase the number of
multiplications, but they would be much faster.
\vspace{5mm}
\noindent

As said, this example was meant to show that there are important applications of polynomial evaluation which can take advantage of a complexity reduction and that there are certainly many other possibilities to further reduce the costs, depending on the particular problem at hand, the model in consideration and the available technology (e.g. availability of storage, of pre-computed tables for finite field mutiplications, etc.).
In particular, this paper has been mainly devoted to the single-point evaluation of polynomials, showing that it is possible to achieve significant complexity reduction with respect to Horner's rule even without any precomputation or storage, especially when the degree of the polynomial is large. In other models, it may be possible to have the powers of $\alpha$ as already given data and to store relatively large binary matrices in order to reduce the number of multiplications in a multi-point evaluation scenario 
or it may be possible to reduce them at the cost of a significant increase of the number of additions. For all these different models, we refer to the vast literature on multi-point evaluation, e.g. \cite{blahut,fedorenko,sarwate}. 

\vspace{2mm}
\noindent
In conclusion, we have proposed some methods to evaluate polynomials in extensions of finite fields that
 have a multiplicative asymptotical complexity $O(\sqrt n)$, much better than $O(n)$, the complexity of standard methods; the constant involved is a function of the field characteristic. 
We have proposed different variants 
  and shown that the choice of an evaluation scheme that uses possibly the smallest number of multiplications follows from a careful analysis of the particular situation
  and might involve the adoption of special tricks dependent on the combination of parameters.
It remains to ascertain whether there exists some evaluation algorithm doing asymptotically better, i.e.
 having a complexity $O(n^t)$ with $t< \frac{1}{2}$.

\section*{Acknowledgments}
The Research was supported in part by the Swiss National Science
Foundation under grant No. 132256.


\end{document}